\documentstyle[12pt]{article}

\catcode`\@=11
\long\def\@makefntext#1{
\protect\noindent \hbox to 3.2pt {\hskip-.9pt
$^{{\ninerm\@thefnmark}}$\hfil}#1\hfill}		%CAN BE USED

\def\@makefnmark{\hbox to 0pt{$^{\@thefnmark}$\hss}}  %ORIGINAL

\def\ps@myheadings{\let\@mkboth\@gobbletwo
\def\@oddhead{\hbox{}
\rightmark\hfil\ninerm\thepage}
\def\@oddfoot{}\def\@evenhead{\ninerm\thepage\hfil
\leftmark\hbox{}}\def\@evenfoot{}
\def\sectionmark##1{}\def\subsectionmark##1{}}

\newcounter{sectionc}\newcounter{subsectionc}\newcounter{subsubsectionc}
\renewcommand{\section}[1] {\vspace*{0.6cm}\addtocounter{sectionc}{1}
\setcounter{subsectionc}{0}\setcounter{subsubsectionc}{0}\noindent
	{\normalsize\bf\thesectionc. #1}\par\vspace*{0.4cm}}
\renewcommand{\subsection}[1] {\vspace*{0.6cm}\addtocounter{subsectionc}{1}
	\setcounter{subsubsectionc}{0}\noindent
	{\normalsize\it\thesectionc.\thesubsectionc. #1}\par\vspace*{0.4cm}}
\renewcommand{\subsubsection}[1]
{\vspace*{0.6cm}\addtocounter{subsubsectionc}{1}
	\noindent {\normalsize\rm\thesectionc.\thesubsectionc.\thesubsubsectionc.
	#1}\par\vspace*{0.4cm}}

\newcounter{appendixc}
\newcounter{subappendixc}[appendixc]
\newcounter{subsubappendixc}[subappendixc]

\renewcommand{\appendix}[1] {\vspace*{0.6cm}
        \refstepcounter{appendixc}
        \setcounter{figure}{0}
        \setcounter{table}{0}
        \setcounter{equation}{0}
        \renewcommand{\thefigure}{\Alph{appendixc}.\arabic{figure}}
        \renewcommand{\thetable}{\Alph{appendixc}.\arabic{table}}
        \renewcommand{\theappendixc}{\Alph{appendixc}}
        \renewcommand{\theequation}{\Alph{appendixc}.\arabic{equation}}
        \noindent{\bf Appendix \theappendixc #1}\par\vspace*{0.4cm}}

\renewenvironment{thebibliography}[1]
	{\begin{list}{\arabic{enumi}.}
	{\usecounter{enumi}\setlength{\parsep}{0pt}
\setlength{\leftmargin 1.25cm}{\rightmargin 0pt}
	 \setlength{\itemsep}{0pt} \settowidth
	{\labelwidth}{#1.}\sloppy}}{\end{list}}

\newcounter{itemlistc}
\newcounter{romanlistc}
\newcounter{alphlistc}
\newcounter{arabiclistc}

\newcommand{\fcaption}[1]{
        \refstepcounter{figure}
        \setbox\@tempboxa = \hbox{\footnotesize Fig.~\thefigure. #1}
        \ifdim \wd\@tempboxa > 6in
           {\begin{center}
        \parbox{6in}{\footnotesize\baselineskip=12pt Fig.~\thefigure. #1}
            \end{center}}
        \else
             {\begin{center}
             {\footnotesize Fig.~\thefigure. #1}
              \end{center}}
        \fi}

\newcommand{\tcaption}[1]{
        \refstepcounter{table}
        \setbox\@tempboxa = \hbox{\footnotesize Table~\thetable. #1}
        \ifdim \wd\@tempboxa > 6in
           {\begin{center}
        \parbox{6in}{\footnotesize\baselineskip=12pt Table~\thetable. #1}
            \end{center}}
        \else
             {\begin{center}
             {\footnotesize Table~\thetable. #1}
              \end{center}}
        \fi}

\def\@citex[#1]#2{\if@filesw\immediate\write\@auxout
	{\string\citation{#2}}\fi
\def\@citea{}\@cite{\@for\@citeb:=#2\do
	{\@citea\def\@citea{,}\@ifundefined
	{b@\@citeb}{{\bf ?}\@warning
	{Citation `\@citeb' on page \thepage \space undefined}}
	{\csname b@\@citeb\endcsname}}}{#1}}

\newif\if@cghi
\def\cite{\@cghitrue\@ifnextchar [{\@tempswatrue
	\@citex}{\@tempswafalse\@citex[]}}
\def\citelow{\@cghifalse\@ifnextchar [{\@tempswatrue
	\@citex}{\@tempswafalse\@citex[]}}
\def\@cite#1#2{{$\null^{#1}$\if@tempswa\typeout
	{IJCGA warning: optional citation argument
	ignored: `#2'} \fi}}

 1
 1
 1

\font\ninerm=cmr9

\begin{document}

\centerline{\normalsize\bf ASPECTS OF GALILEAN AND RELATIVISTIC PARTICLE}
\baselineskip=16pt
\centerline{\normalsize\bf MECHANICS WITH DIRAC'S CONSTRAINTS}
\baselineskip=15pt

%\vfill
\vspace*{0.6cm}
\centerline{\footnotesize Luca Lusanna}
\baselineskip=13pt
\centerline{\footnotesize\it Sezione INFN di Firenze}
\baselineskip=12pt
\centerline{\footnotesize\it Largo E.Fermi 2 (Arcetri)}
\baselineskip=12pt
\centerline{\footnotesize\it 50125 Firenze, Italy}
\centerline{\footnotesize E-mail: lusanna@fi.infn.it}

%\vfill
\vspace*{0.9cm}

Relevant physical models are described by singular Lagrangians, so that
their Hamiltonian description is based on the Dirac theory of
constraints\cite{diraca}. The qualitative aspects of this theory are now
understood\cite{lusa}, in particular the role of the Shanmugadhasan
canonical transformation\cite{shan} in the determination of a canonical basis
of Dirac's observables allowing the elimination of gauge degrees of
freedom from the classical description of physical systems\cite{lusb}. This
programme was initiated by Dirac\cite{diracb} for the electromagnetic field
with charged fermions. Now Dirac's observables for Yang-Mills theory with
fermions (whose typical application is QCD) have been found\cite{lusc} in
suitable function spaces where the Gribov ambiguity is absent. Also the ones
for the Abelian Higgs model are known\cite{lusd} and those for the $SU(2)
\times U(1)$ electroweak theory with fermions are going to be found\cite{luse}
with the same method working for the Abelian case. The main task along these
lines will now be the search of Dirac's observables for tetrad gravity in the
case of asymptotically flat 3-manifolds.

The philosophy behind this approach is ``first reduce, then quantize": this
requires a global symplectic separation of the physical variables from the
gauge ones so that the role of differential geometry applied to smooth
field configurations is dominating, in contrast with the standard approach
of ``first quantizing, then reducing", where, in the case of gauge field
theory, the reduction process takes place on distributional field
configurations, which dominate in quantum measures. This global separation has
been accomplished till now, at least at a heuristic level, and one is
going to have a classical (pseudoclassical for the fermion
variables\cite{casal}) basis for the physical description of the
$SU(3)\times SU(2)\times U(1)$ standard model; instead, with tetrad gravity one
expects to get local results in accord with the interpretational problems of
general relativity with its invariance under diffeomorphisms. However, the
price for having only physical degrees of freedom is the non-local (and in
general non-polynomial) nature of the physical Hamiltonians and
Lagrangians, as already known from Dirac's paper on the electromagnetic
field\cite{diracb} due to the Coulomb self-interaction of the fermion fields.
Two obstacles appear immediately: 1) the lack of manifest Lorentz covariance
of the Hamiltonian formalism, which requires the choice of a 3+1 splitting of
Minkowski space-time; 2) the inapplicability of the standard methods of
regularization and renormalization due to the non-locality (and
non-polynomiality) of the interactions and the failure of the power
counting rule.

\bigskip
Talk given at the Conference ``Theories of Fundamental Interactions",
Maynooth (Ireland), May 1995.

\vfill\eject

The problem of Lorentz covariance is present in all schemes of reduction,
either before or after quantization; it is impossible to eliminate all the
gauge degrees of freedom in a Lorentz covariant way. As shown in
Ref.[6], there is one way to obtain the minimal breaking of Lorentz
covariance, i.e. to reformulate classical field theory on a family of
arbitrary space-like hypersurfaces\cite{diraca} foliating Minkowski
space-time and then to restrict ourselves to the family of hyperplanes
orthogonal to the total four-momentum $P_{\mu}$ of the field configuration,
when it is time-like; only field configurations in irreducible representations
of the Poincar\'e group are considered: they satisfy suitable boundary
conditions implying that the ten functionals defining the Poincar\'e
generators are finite. In this way only three physical degrees of freedom,
describing the canonical center-of-mass 3-position of the overall isolated
system, break Lorentz covariance, while all the field variables are either
Lorentz scalars or Wigner spin-1 3-vectors transforming under Wigner
rotations. This method is based on canonical realizations of the Poincar\'e
group on spaces of functions on phase spaces and one has the transposition at
the canonical level of the techniques used to study the irreducible
representations of the Poincar\'e group and the relativistic wave equations.

The problem of the inapplicability of the power counting rule is connected
with the unsolved problem of regularizing the Coulomb gauge in QED, even if
there is no theorem implying its impossibility. In the Yang-Mills case one
gets not only non-local but also non-polynomial self-interactions, so that one
cannot use these results for canonical quantization at this stage. On the
other hand, all the standard techniques of regularization fail in the case
of general relativity. If it will be possible to solve 13 of the 14 first
class constraints of tetrad gravity, the final form of the super-Hamiltonian
constraint will be non-local and non-polynomial in the graviton Dirac
observables as is the physical Hamiltonian of Yang-Mills theory. One could then
couple tetrad gravity to the standard model and try again to find the
Shanmugadhasan canonical transformation, arriving at a result in which all
the interactions are put on the same level with a final form of the
non-locality and non-polynomiality. Every advance in understanding the
quantization of the system would apply to all the interactions, and
moreover one would have a framework for trying to find a suitable
definition of elementary particles fitting with particle physics when
restricted to Minkowski space-time.

At this stage only one tool is emerging: the problem of the center-of-mass
extended relativistic systems in irreducible representations of the Poincar\'e
group with $P^2 > 0$, $W^2=-P^2{\vec {\bar S}}^2\not= 0$ (they are dense in
the set of all allowed field configurations) identifies a finite world-tube of
non-covariance of the canonical center-of-mass, whose radius $\rho =\sqrt{-W^2}
/P^2=|\, {\vec {\bar S}}\, |\, /\sqrt{P^2}$ identifies a classical intrinsic
unit of length, which can be used as a ultraviolet cutoff at the quantum
level in the spirit of Dirac and Yukawa. As mentioned in Ref.[6], the distances
corresponding to the interior of the world-tube are connected with problems
coming from both quantum theory and general relativity: 1) pair production
happens when trying to localize particles at these distances; 2) relativistic
bodies with a material radius less than $\rho$ cannot have the classical
energy density definite positive everywhere in every reference frame and the
peripheral rotation velocity may be higher than the velocity of light.
Therefore, the world-tube is the flat remnant of the energy  conditions of
general relativity; in this theory
the radius $\rho$ is defined in terms of the asymptotic
Poincar\'e group existing in the case of asymptotically flat 3-manifolds.

However, it is not clear how to use this cutoff in a constructive way in
canonical quantization of non-local and non-polynomial theories
(forgetting at this level ordering problems in the physical Hamiltonian).
To clarify the situation, a preliminary step would be to find a
center-of-mass and relative variable decomposition of a field
configuration, in analogy to what has already been done for
two\cite{longhi} or
N\cite{lusf} scalar particles and for the Nambu string\cite{colomo}. In
parallel to this problem, which is now under investigation for the Klein-Gordon
field, one should need a reformulation of classical field theory and of its
Cauchy problem in this kinematical framework.

Actually, the standard Fock space has asymptotic states defined as tensor
products of free one-particle states and the standard perturbative expansions
correspond to the propagation of off-shell free intermediate particles. Now,
in a tensor product there is no restriction on the relative-time correlations
among the free particles: one asymptotic free particle may be in the
absolute future of another one, since there is no mechanism inhibiting such a
possibility. Whereas this fact may be irrelevant for scattering processes in
the
S-matrix approach, it becomes a problem for relativistic bound states: it is
known that, in general, the 2-body Bethe-Salpeter equation\cite{saz} has
spurious solutions, which are excitations in relative energy, the variable
conjugate to relative time. Therefore, one needs a reformulation of quantum
field theory and of its asymptotic states with the problem of relative
times and energies of the asymptotic particles under control. The natural
framework is again the formulation of classical field theory on
space-like hypersurfaces, which is the classical basis of the
Tomonaga-Schwinger formulation of quantum field theory (whose asymptotic
states do not seem to have been defined). If, for $P^2 > 0$, one restricts
oneself to space-like hyperplanes orthogonal to the total momentum $P_{\mu}$,
one obtains a covariant formulation of the instant form of dynamics of
Dirac\cite{diracc} (which could be called the ``rest-frame form") with the
Lorentz-scalar rest-frame time T as the time variable. In this way one
obtains a one-time (T) theory with a well defined Hamiltonian (like in the
Newton case) for the reduced problem, after a separation of the free
non-covariant canonical center-of-mass motion, associated with the
physical system (fields and/or particles) lying on the hyperplane. The problem
with relative times and energies disappears by construction; when the
center-of-mass of a field configuration will be under control, one will develop
a ``rest-frame quantum field theory", whose asymptotic states will be free
particles on the hyperplane. This will imply a perturbative expansion in which
the only off-shell propagation will involve the overall system (the inverse
propagator will be the mass-shell constraint for the isolated system) and not
the single particles. It is in this framework that the previous ultraviolet
cutoff becomes meaningful. Moreover, the resulting bound state equations will
be free from spurious solutions by construction and, hopefully, one will have
a coherent starting point for the introduction of bound states among the
asymptotic states.

To arrive at this description, a revisitation of classical relativistic
mechanics for 2 and N scalar particles was needed; the addition of spin
degrees of freedom with Grassmann variables\cite{crater}  is only a technical
complication at this point. The two main problems which slowed down the
development of relativistic mechanics were: a) the
No-Interaction-Theorem\cite{nit,nita} [see Ref.[18] for a review];
b) the many definitions  of relativistic center-of-mass position [see
Ref.[10] for reviews]. While the latter problem is due to the
Lorentz signature of Minkowski space-time, the former is connected with the
multi-time description of particle dynamics. The No-Interaction-Theorem
was discovered in the relativistic context, and thought
to be connected with the
Lorentz signature, which requires 4-vector configuration variables
$q^{\mu}_i$ to describe world-lines in a Lorentz covariant way. Predictive
mechanics\cite{bel} used as configuration variables ${\vec q}_i(q^o_i)$, wrote
Newton-like equations $m_id^2{\vec q}_i(q^o_i)/dq^{o\, 2}_i={\vec F}_i$ and
found the Currie-Hill conditions\cite{ch} which had to be satisfied by the
functions ${\vec F}_i$ to be admissible relativistic forces (they are the
necessary and sufficient conditions for having a Lorentz invariant
dynamics\cite{bel}). In this description it is emphasized that one may
reparametrize each world-line independently from the others, like it happens
in the Fokker-Tetrode-Feynman-Wheeler actions, which, however, give
integro-differential equations of motion (it is conjectured that they admit a
subset of predictive solutions\cite{gaida} with Newtonian Cauchy data if one
adds a selection rule of the type: choose those solutions analytic in the
coupling constants (or in the inverse of the light velocity) such that
turning off the coupling constant only free motion survive).

Since the Currie-Hill conditions  are too complicated to be solved, a
Hamiltonian formulation was developed under the hypothesis that the
configuration variables ${\vec q}_i$ coincide with the canonical ones ${\vec
x}_i$, ${\vec q}_i={\vec x}_i$, in the instant form of dynamics. Then, the
requirements that the Lorentz boosts can be implemented as canonical
transformations and that the transformation ${\vec q}_i, {\vec v}_i\mapsto
{\vec q}_i={\vec x}_i, {\vec p}_i$ is non singular, implied that only free
motion is allowed: this is the original form of the
No-Interaction-Theorem\cite{nit}. Another form\cite{nita} makes the
hypothesis that rotations and space-time translations are implemented as
canonical transformations with constant generators ${\vec J}$, $\vec P$,
$P^o=H$ (the Hamiltonian): again only free motion is allowed, implying that a
predictive Hamiltonian H, and therefore a predictive Lagrangian, does not
exist.

In an attempt to understand relativistic predictive mechanics, where the
absence of an absolute definition of time, due to Lorentz signature, makes
the description so complicated, nonrelativistic Newton equations were
reformulated as multi-time equations\cite{pons} by rescaling the time
parameter t in their solutions independently for each particle, ${\vec q}_i(t)
\mapsto {\vec q}_i(t_i)$. The equations of motion become $m_id^2{\vec
q}_i(t_i)/
dt^2_i={\vec F}_i({\vec q}_k(t_k),d{\vec q}_k(t_k)/dt_k,t_k)$, so that the
predictive conditions, replacing the Currie-Hill conditions, are now
$d{\vec F}_i/dt_j=0$, $j\not= i$. It turned out that the nonrelativistic
predictive Lagrangian does not exist like in the relativistic case (only
line actions associated with lines in the parametric multi-time space can
be defined). By shifting to the first-order formalism and then by solving a
Pfaff problem connected with the multi-time generalization of the Lie-K\"onig
theorem, multi-time Hamiltonian formulations of Galilean predictive mechanics
for N particles were found, with N Hamiltonians $H_i$ (one for each time)
satisfying $\partial H_i/\partial t_j\, -\, \partial H_j/\partial t_i\, +
\lbrace H_i,H_j\rbrace =0$. The requirement of having Galilei transformations
implemented as canonical transformations, selects one (or very few in general)
of these multi-time symplectic structures (see Ref.[24,23] for a
discussion of the conditions for having uniqueness in the selection). The
main point is that the canonical coordinates are functions of all the times
simultaneously, ${\vec x}_i={\vec x}_i(t_1,..,t_N)$, with the only exception
of the free case in which ${\vec q}_i(t_i)={\vec x}_i(t_i)$; but always one has
${\vec q}_i(t)={\vec x}_i(t,..,t)$ at equal times. This shows that the
No-Interaction-Theorem is independent of the Lorentz signature, that it arises
from the multi-time description of dynamics if one requires ${\vec q}_i={\vec
x}_i$, and, moreover, that, by enlarging the phase space with the addition of
the canonical  pairs $t_i$, $E_i$, $\lbrace t_i,E_j\rbrace =-\delta_{ij}$, one
obtains a Hamiltonian formulation with N first class constraints $\chi_i=
E_i-H_i\approx 0$, $\lbrace \chi_i,\chi_j\rbrace =0$. This implies the
existence
of singular Lagrangians with configuration variables $t_i(\tau )$, ${\vec x}_i
(\tau )$, generating these constraints; if one puts $t_1(\tau )=\cdots =t_N
(\tau )=t(\tau )$ inside these Lagrangians, one recovers the parametrized form
of Newton Lagrangians with the configuration variables ${\vec q}_i(\tau )=
{\vec x}_i(\tau )$, $t(\tau )$ and one first class constraint $E-H\approx 0$,
$E=\sum_iE_i$, $H=\sum_iH_i{|}_{t_1=\cdots =t_n=t}$, (H is the Newtonian
Hamiltonian). In Ref.[23] there is an explicit form of the constraints
for N=2 and the form of the singular Lagrangian for the 2-time harmonic
oscillator.

This clarification of the meaning of the theorem allowed to understand the
connection of relativistic predictive mechanics with the 2-body
Droz Vincent-Todorov-Komar model\cite{dv,todo,komar} based on two first
class constraints $\chi_i=p_i^2-m_i^2+V(r^2_{\perp})\approx 0$, $\lbrace
\chi_1,\chi_2\rbrace =0$, $r^{\mu}_{\perp}=(\eta^{\mu\nu}-P^{\mu}P^{\nu}/
P^2)r_{\nu}$, $r^{\mu}=x_1^{\mu}-x_2^{\mu}$, $P^{\mu}=p_1^{\mu}+p_2^{\mu}$
[see Ref.[28] for previous Lagrangian models implying second class
constraints]. Since the canonical Hamiltonian is zero and since with each first
class constraint
there is associated an arbitrary Dirac multiplier $\lambda_i(\tau
)$, one can define a 2-time theory\cite{lusc} by defining $d\tau_i=\lambda_i
(\tau )d\tau$, $x^{\mu}_i(\tau ), p^{\mu}_i(\tau )\mapsto x_i^{\mu}(\tau_1,
\tau_2), p_i^{\mu}(\tau_1,\tau_2)$, and with $\chi_i$ as the Hamiltonians for
the evolution in the $\tau_i$'s. While at the nonrelativistic level [$\chi_i=
E_i-{\vec p}_i^2/2m_i + V({\vec \rho}^2)/2m_i\approx 0$, $\vec \rho =\vec r-
t_R\vec P/(m_1+m_2)$, $t_R=t_1-t_2$, $\vec P={\vec p}_1+{\vec p}_2$, $\vec r=
{\vec x}_1-{\vec x}_2$] one has $d\tau_i=-dt_i$, at the relativistic level the
situation is more complex, because one has to make a 3+1 splitting of
Minkowski space-time or a choice of which Dirac form of dynamics to use;
moreover, one has to classify the motions according to the value of the
Poincar\'e Casimir $P^2$ (the constraint manifold is a stratified manifold
due to the various kinds of Poincar\'e orbits). Restricting ourselves to the
main stratum $P^2 > 0$, the natural choice would be the instant form $x^o=
const.$, which however is not covariant. However, in the model one can recover
the predictive positions $q^{\mu}_i(\tau_i)$ by solving Droz Vincent's
equations\cite{dv} $\lbrace q^{\mu}_i,\chi_j\rbrace =0$, $i\not= j$, with the
boundary conditions $q^{\mu}_i=x^{\mu}_i$ on the hypersurface $P\cdot r=0$
(equal times in the rest frame). This suggests the existence of a covariant
rest-frame instant form of dynamics defined by a foliation whose leaves are
space-like hyperplanes labelled by a rest-frame Lorentz-scalar time
$T=P\cdot x/\sqrt{P^2}$ with $x^{\mu}$ some center-of-mass coordinate; in the
rest-frame instant form all the particles have the same time of the surface
and the 1-time evolution should be governed by only one first class constraint:
for two particles it is $\chi =\chi_1+\chi_2$, the constraint determining
the mass spectrum of the isolated system; instead $\chi_1-\chi_2$ determines
$P\cdot q$ ($q^{\mu}$ is the relative momentum) and by adding $P\cdot r
\approx 0$ as a gauge-fixing one obtains a pair of second class constraints
implying the reduction of the dynamics to the rest-frame instant form. In
Ref.[10] there is a complete study, both at the classical and at
the quantum level, of the 2-body model and the definition of a series of
canonical transformations ($\eta =sign\, P^o$):
\begin{eqnarray}
x^{\mu}_i,\, p_i^{\mu}\, &\mapsto & \nonumber\\
&\mapsto & x^{\mu}={1\over 2}(x_1^{\mu}+x_2^{\mu}),
P^{\mu}=p_1^{\mu}+p_2^{\mu}, r^{\mu}=x_1^{\mu}-x_2^{\mu}, q^{\mu}={1\over 2}
(p_1^{\mu}-p_2^{\mu})\mapsto \nonumber\\
&\mapsto & {\hat x}^{\mu}=x^{\mu}+{{m_1^2-m_2^2}\over {2P^2}}(\eta^{\mu\nu}-
2P^{\mu}P^{\nu}/P^2)r_{\nu},\, P^{\mu},\nonumber\\
&{}& r^{\mu},\, {\hat q}^{\mu}=q^{\mu}-
{{P^{\mu}}\over {2P^2}}(m_1^2-m_2^2)\, \mapsto \nonumber\\
&\mapsto &{\tilde x}^{\mu}={\hat x}^{\mu}+{1\over 2}\epsilon^A_{\mu}(u(P))
\eta_{AB}{{\partial \epsilon^B_{\rho}(u(P))}\over {\partial P_{\mu}}}
S^{\rho\nu},\, P^{\mu}, \rho^a=\epsilon^a_{\mu}(u(P))r^{\mu},\nonumber\\
&{}& \pi^a=
\epsilon^a_{\mu}(u(P)){\hat q}^{\mu},\, T_R=\epsilon^o_{\mu}(u(P))r^{\mu}=
P\cdot r/\eta \sqrt{P^2},\nonumber\\
&{}& {\hat \epsilon}_R=\epsilon^o_{\mu}(u(P)){\hat
q}^{\mu}=P\cdot {\hat q}/\eta \sqrt{P^2} \mapsto \nonumber\\
&\mapsto &
\vec z=\eta \sqrt{P^2}({\vec {\tilde x}}-{{\vec P}\over {P^o}}{\tilde x}^o),\,
\vec k=\vec u(P)=\vec P/\eta \sqrt{P^2},\nonumber\\
&{}& T=P\cdot \tilde x/\eta \sqrt{P^2}=
P\cdot \hat x/\eta \sqrt{P^2},\, \epsilon =\eta \sqrt{P^2},\, \vec \rho ,\,
\vec \pi ,\,T_R,\, {\hat \epsilon}_R\nonumber\\
\end{eqnarray}
where $\epsilon^A_{\mu}(u(P))=L^A{}_{\mu}(P_R,P)$ [$\epsilon^o_{\mu}(u(P))=
u_{\mu}(P)=P_{\mu}/\eta \sqrt{P^2}$] are the rows of the standard Wigner boost
to the rest frame for time-like orbits [$L^{\mu}{}_{\nu}(P_R,P)P^{\nu}=P_R
^{\mu}=(\eta \sqrt{P^2};\vec 0)$] and $J^{\mu\nu}=\sum_i(x_i^{\mu}p_i^{\nu}-
x_i^{\nu}p_i^{\mu})={\hat x}^{\mu}P^{\nu}-{\hat x}^{\nu}P^{\mu}+S^{\mu\nu}=
{\tilde x}^{\mu}P^{\nu}-{\tilde x}^{\nu}P^{\mu}+{\tilde S}^{\mu\nu}$ with
$S^{\mu\nu}=r^{\mu}{\hat q}^{\nu}-r^{\nu}{\hat q}^{\mu}$, ${\tilde S}^{io}=
\rho^i{\hat \epsilon}_R-T_R\pi^i={\tilde S}^{ij}P_j/(P^o+\eta \sqrt{P^2})$,
${\tilde S}^{ij}=\rho^i\pi^j-\rho^j\pi^i$. $\vec \rho$, $\vec \pi$ are Wigner
spin-1 3-vectors transforming under Wigner rotations and ${\tilde x}^{\mu}$,
$\vec z$ do not transform covariantly under Lorentz boosts. The two constraints
may now be written as ${\hat \epsilon}_R\approx 0$, $[\epsilon^2-M^2_{+}(\vec
\rho ,\vec \pi )][\epsilon^2-M^2_{-}(\vec \rho \vec \pi )]\approx 0$ with
$M_{\pm}(\vec \rho ,\vec \pi )=\sqrt{m_1^2+{\vec \pi}^2+V(-{\vec \rho}^2)}\pm
\sqrt{m_2^2+{\vec \pi}^2+V(-{\vec \rho}^2)}$. While the second one determines
the four branches of the mass spectrum, with the conjugate gauge variable being
the rest-frame time T (the natural choice of a clock for the global evolution
would be obtained with the gauge-fixing $T-\tau \approx 0$), the first one
implies that the relative time $T_R$ is a gauge variable in the sense that one
observer has the freedom to describe the two particles either at equal
rest-frame
time ($T_R\approx 0$) or with every time delay he wishes. The last canonical
transformation defines a quasi-Shanmugadhasan canonical basis containing
${\hat \epsilon}_R$ ($\approx 0$), but not the mass-shell constraint: there
are four disjoint branches with total invariant mass (the effective
Hamiltonian for the evolution in T) $\pm M_{+}$, $\pm M_{-}$, so that one
should get four final Shanmugadhasan canonical bases, but only if the
dynamics is Liouville integrable (the final Dirac observables should include,
besides $\vec z, \vec k$, also the associated angle-action variables replacing
$\vec \rho ,\vec \pi$). In Ref.[11] the same construction was attempted
for N free scalar particles ($\chi_i=p_i^2-m^2_i\approx 0$) with the attempt to
replace N-1 combinations of the constraints with the vanishing of N-1 relative
energies ${\hat \epsilon}_{Ra}\approx 0$, a=1,..,N-1, but only partial results
were obtained due to the non-linearity of the canonical transformations. One
could not identify a parametrization of the $2^N$ branches of the mass spectrum
and have a kinematical control of the N-time description of the particles due
to algebraic complications.

Let us remark that in the sequence of canonical transformations for the 2-body
case various definitions of center-of-mass position appeared, which reflect
the complexity of this concept at the relativistic level due to Lorentz
signature. While the predictive configuration coordinates $q^{\mu}_i(\tau )$
(solutions of the Droz Vincent equations) are non-canonical ($\lbrace q^{\mu}
_i,q^{\nu}_j\rbrace \not= 0$) 4-vectors describing the geometrical world-lines,
the canonical coordinates $x^{\mu}_i(\tau )$ [or $x^{\mu}_i(\tau_1,\tau_2)$ in
the multi-time approach] are 4-vectors which couple minimally to external
fields.
The naive center-of-mass $x^{\mu}+{1\over 2}(x_1^{\mu}+x_2^{\mu})$ is canonical
and a 4-vector, but does not have a free motion [it has a ``classical
zitterbewegung" due to the action-at-a-distance potential $V(r^2_{\perp})$].
The  variables ${\tilde x}^{\mu}$ and $\vec z/\eta \sqrt{P^2}$ are canonical
but not covariant under Lorentz boosts: ${\tilde x}^{\mu}$ has free motion,
while $\vec z/\eta \sqrt{P^2}={\vec {\tilde x}}-{{\vec P}\over {P^o}}{\tilde
x}^o=-{{\vec K}\over {P^o}}+{ { {\vec {\tilde S}}\times \vec P}\over
{P^o(P^o+\eta \sqrt{P^2})} }$ [$K^i=J^{oi}$] is a frozen Jacobi data;
$\vec z/\eta \sqrt{P^2}$ is the classical analogue of the
Pryce-Newton-Wigner\cite{nw} 3-position operator (it is also called the
center of spin) and it can be shown\cite{pauri} that it is built entirely in
terms of the Poincar\'e generators (see the theory of the canonical
realizations of the Poincar\'e group\cite{pp}); since it is not a 4-vector,
it defines a different world-line in every boosted frame, so that it is better
to denote it with a frame index, ${\vec z}_F/\eta \sqrt{P^2}$ (${\tilde x}
^{\mu}_F$). In Ref.[30] it is shown that using only the generators of
the Poincar\'e group one can obtain two other definitions of center-of-mass
3-positions at ${\tilde x}^o=0$: a) Fokker's center of inertia\cite{fok}
$\vec Y=-{{\vec K}\over {P^o}}+{ { {\vec {\tilde S}}\times \vec P}\over
{P^o\eta \sqrt{P^2}} }$, which is not canonical ($\lbrace Y^i,Y^j\rbrace \not=
0$) but defines a 4-vector by construction ($\vec Y={\vec z}_{F=R}/\eta \sqrt
{P^2}$ in the rest frame at ${\tilde x}^o=0$; then it is defined in other
frames by applying to it the Lorentz transformation connecting the frame to
the rest frame); b) M\"oller center-of-mass\cite{moll} ${\vec R}_F=-{{\vec K}
\over {P^o}}$, which is neither canonical nor covariant (it is defined
starting from $\sum_i{{p^o_i}\over {P^o}}{\vec x}_i$ at $x^o_i=0$ by replacing
the Newtonian masses with the energies). In the rest frame ${\vec z}_{F=R}/
\eta \sqrt{P^2}$, $\vec Y$, ${\vec R}_{F=R}$ coincide and in every frame they
have the same velocity $\vec P/P^o$. While $x^{\mu}$ behaves like the Dirac
position of a fermion, ${\tilde x}^{\mu}$ is similar to the
Foldy-Wouthuysen mean position\cite{fw} except for its non-covariance; in the
case of certain pole-dipole systems\cite{pauri}, there is a 3-position, built
with the Poincar\'e generators and an extra variable, which is both canonical
and covariant (it is a generalization of the mean position), but it does not
seem to exist for extended systems; however this is still an open problem.

If, in any given reference frame, one draws all the pseudo-world-lines ${\vec
z}_F/\eta \sqrt{P^2}$, ${\vec R}_F$ associated with all possible frames, one
obtains a world-tube around the Fokker center of inertia, whose invariant
radius\cite{moll,pauri}
is $\rho =\sqrt{-W^2}/P^2=|\, {\vec {\tilde S}}\,| /\sqrt{P^2}$ if
$P^2 > 0$ (${\vec {\tilde S}}$ is the rest-frame Thomas spin). This world-tube,
as already anticipated, has the following remarkable properties: a) a measure
of the canonical 3-position ${\vec z}_F/\eta \sqrt{P^2}$ would be
frame-dependent, while a measure of the classical Fokker center of inertia
would not have a quantum counterpart because $[{\hat Y}^i,{\hat Y}^j]\not= 0$;
b) the criticism to the classical theory from the quantum point of view
based on pair production would apply to distances less than $\rho$ (at the
quantum level it would be the Compton wavelength of the isolated system
multiplied by the value of its total spin; also the quantum zitterbewegung
of the Newton-Wigner 3-position operator would be inside the world-tube);
c) a material body with radius less than $\rho$ could have a peripheral
rotation velocity higher than the light velocity and would not have the
classical energy density definite positive everywhere in every reference
frame\cite{moll} (the world-tube is a remnant of the energy conditions of
general relativity in flat Minkowski space-time). Therefore, there is a
conceptual problem about either the classical or quantum localization of
the center-of-mass of an extended relativistic system. This problem with the
theory of relativistic measurements suggests to implement Heisenberg
indetermination relations for the center-of-mass with $\triangle x^i \geq
\rho$,
to abandon the requirement of
self-adjointness of $\vec x=\vec z/\eta \sqrt{P^2}$ and to
accept only wave packets constant inside the world-tube (democracy of all
reference frames) and with power tails so as to avoid Hegerfeldt's
theorems\cite{heger}: the wave packets would spread with a velocity less than
the velocity of light. After all on one side the center-of-mass of a field
configuration is a concept like the wave function of the universe, while the
actual observability of the center-of-mass of a system of N relativistic
particles presumably will never imply its localization inside the world-tube.

After all these preliminaries, let us come back to systems of N relativistic
particles. To overcome the quoted algebraic difficulties with the N-time
description with N first class constraints and to be able to treat
simultaneously N charged particles and the electromagnetic field [one does not
know how to evaluate the Poisson bracket of the minimally coupled particle
constraints ${(p_i(\tau )-e_iA(x_i(\tau )))}^2-m_i^2\approx 0$ and of the
primary field constraints $\pi^o(\vec z,z^o)\approx 0$, lacking a covariant
notion of equal times], it seems that the only way out is to reformulate the
theory on space-like hypersurfaces\cite{diraca}. But now all the particles are
simultaneous with respect to the hypersurfaces, the coordinates of whose
points are $z^{\mu}(\tau ,\vec \sigma )$, and are labelled by only three
curvilinear coordinates $\vec \sigma ={\vec \eta}_i(\tau )$. This conceptually
implies the addition of N-1 gauge-fixings for the relative times, so that
only one first class constraint giving the mass spectrum of the whole system is
expected. It also implies that for each particle one has to choose one of the
two branches of its mass-hyperboloid, i.e. the sign of its energy, $\alpha_i=
sign\, p_i^o$. As shown in Refs.[36,6,37], one can restrict the
infinite number of first class constraints ${\cal H}_{\mu}(\tau ,\vec \sigma )
\approx 0$, implying the independence of the description from the choice of
the hypersurface with coordinates $z^{\mu}(\tau ,\vec \sigma )$, to only
four first class constraints by adding suitable gauge-fixings, which reduce the
hypersurfaces to the family of hyperplanes orthogonal to the total
four-momentum
of the system (it could be called the Wigner foliation). In this way only a
non-canonical 4-vector $x^{\mu}(\tau )$ ($\lbrace x^{\mu},x^{\nu}\rbrace
\not= 0$) is left of all the $z^{\mu}(\tau ,\vec \sigma )$'s and a canonical
non-covariant ${\tilde x}^{\mu}(\tau )$ may be built from it: one has obtained
the rest-frame instant form with time $T=P\cdot x/\eta \sqrt{P^2}$ and with the
world-tube of center-of-mass 3-positions emerging naturally. The components
$A_o(\tau ,\vec \sigma )$, $\vec A(\tau ,\vec \sigma )$ of the electromagnetic
potential become a scalar and a Wigner spin-1 3-vector and the reduction to
the transverse Dirac observables ${\vec A}_{\perp}$, ${\vec E}_{\perp}$, of
the radiation field may be performed
covariantly. Moreover the Coulomb potential
between the charged particles (endowed with Grassmann-valued electric charges
to avoid classical self-energies) emerges naturally from the solution of the
Gauss law and from the identification of the Dirac observables for the particle
momenta. One of the four constraints determines the mass spectrum of the
system (the resulting invariant mass is the Hamiltonian for the evolution in
T), while the other three constraints imply the vanishing of the total
3-momentum on the hyperplane, which is orthogonal to the total 4-momentum.
The details of the construction will be given elsewhere\cite{lusg}. Let us
remark that the rest-frame instant form allows us
to give a Lorentz-scalar 1-time
description of the dynamics with a Lorentz-scalar Hamiltonian for all the
configurations of the system with $P^2 > 0$ in a way similar to Newtonian
physics, so that, for instance, one can develop a relativistic statistical
mechanics (it is possible to define a perfect gas of relativistic oscillators).
Moreover, it will be the basis for defining the rest-frame field theory,
based on the center-of-mass and relative variable decomposition of classical
field configurations; it is hoped that the quantization of these relative
variables will allow the definition of the asymptotic states of the
Tomonaga-Schwinger formulation of quantum field theory, a consistent
utilization
of the radius of the world-tube as an ultraviolet cutoff and a definition of
bound state equations free of the spurious solutions of the Bethe-Salpeter
equation.

\end{document}